\documentclass[%
 reprint,
showpacs,
 amsmath,amssymb,
 aps,
 pra,
longbibliography,
]{revtex4-1}

\usepackage[utf8]{inputenc}	
\usepackage[T1]{fontenc}	
\usepackage{graphicx}		
\usepackage{color}
\usepackage{pdfpages}
\usepackage[english]{babel}
\usepackage{hyperref}
\usepackage{url}
\usepackage{todonotes}
\usepackage{lipsum}
\usepackage{bm}

\usepackage{braket}

\renewcommand{\vec}[1]{\bm{#1}}

\usepackage[position=top,caption=false]{subfig}

\let\originaleqref\eqref
\renewcommand{\eqref}{Eq.~\originaleqref}

\newcommand{\fref}[1]{\figurename~\ref{#1}}

\newcommand{\e}[1]{\mathrm{e}^{#1}}

\newcommand{\Tr}[1]{\mathrm{Tr}\left(#1\right)}

\newcommand{\pdiff}[2]{\frac{\partial #1}{\partial #2}}
\newcommand{\Int}[4]{\int_{#1}^{#2}d#3\,#4}

\newcommand{\C}[0]{\hat{c}}
\newcommand{\Cd}[0]{\hat{c}^\dagger}

\newcommand{\tp}[0]{{t^\prime}}

\newcommand{\E}[1]{\mathbb{E}\left[#1\right]}
\newcommand{\Fet}[0]{F^{(1)}_\Phi}

\newcommand{\Cl}[0]{C(\tau)}
\newcommand{\xPhi}[0]{\mathcal{X}_{\Phi}}

\newcommand{\Sw}[0]{S_{\Phi}(\omega)}

\renewcommand{\L}[0]{\mathcal{L}}
\newcommand{\eLt}[1]{\e{\mathcal{L}#1}}
\newcommand{\rst}[0]{\rho_\mathrm{st}}
\newcommand{\sx}[0]{\hat{\sigma}_\mathrm{x}}
\newcommand{\sy}[0]{\hat{\sigma}_\mathrm{y}}

\renewcommand{\sp}[0]{\hat{\sigma}_\mathrm{+}}
\newcommand{\sm}[0]{\hat{\sigma}_\mathrm{-}}

\newcommand{\Iet}[0]{\mathcal{I}^{(1)}}
\newcommand{\Ito}[0]{\mathcal{I}^{(2)}}

\graphicspath{
{./Plots/}
}

\begin{document}
\title{Bayesian parameter estimation by continuous homodyne detection}
\author{Alexander Holm Kiilerich}
\email{kiilerich@phys.au.dk}
\author{Klaus Mølmer}

\date{\today}
\affiliation{Department of Physics and Astronomy, Aarhus University, Ny Munkegade 120, DK 8000 Aarhus C. Denmark}
\date{\today}

\bigskip

\begin{abstract}
We simulate the process of continuous homodyne detection of the radiative emission from a quantum system, and we investigate how a Bayesian analysis can be employed to determine unknown parameters that govern the system evolution. Measurement backaction quenches the system dynamics at all times and we show that the ensuing transient evolution is more sensitive to system parameters than the steady state of the system. The parameter sensitivity can be quantified by the Fisher information, and we investigate numerically and analytically how the temporal noise correlations in the measurement signal contribute to the ultimate sensitivity limit of homodyne detection.
\end{abstract}

\maketitle
\noindent


\section{Introduction}

Single quantum systems such as atoms, light fields and mechanical oscillators have found wide applications in quantum enhanced metrology and as sensitive probes of weak perturbations \cite{PhysRevLett.96.010401,Giovannetti19112004}. In many experiments, precision is acquired by continuously probing a single system over time rather than by employing single measurements on many identical probe systems. For continuous probing, we may expect that the measurement precision improves with the total measurement duration $T$, and if measurement outcomes at different times $(t,t')$ are  uncorrelated for $|t-t'| \geq \tau$, we may compare the situation with that of $N\sim T/\tau$ independent measurements, and hence expect an estimation error scaling as $1/\sqrt{T}$.

A quantitative analysis of the estimation precision of an experiment relies strongly on the kind of measurement performed on the system. Apart from the dependence of the different outcome probabilities on the quantum state via Born's rule, we must take into account the backaction of each measurement on the state of the system, as this governs the subsequent evolution and thus affects future outcome probabilities. In the case of counting of photons emitted from an atomic system, the detection of a photon at a random time is accompanied by a quantum jump where the atom is reset to its ground state, and the time intervals between emission events are thus governed by the (same) transient excited state probability function. Unlike the mean fluorescence intensity, which is given by the steady state excitation of the atom and saturates for strong driving, the transient evolution shows oscillations at the driving Rabi frequency, and hence the counting measurements and, notably, their backaction, lead to better resolution of large Rabi frequencies and other interaction parameters \cite{PhysRevA.89.052110,PhysRevA.91.012119}.

\begin{figure}
\subfloat[\label{fig:setup}]{
\includegraphics[trim=0 0 0 0,width=0.9\columnwidth]{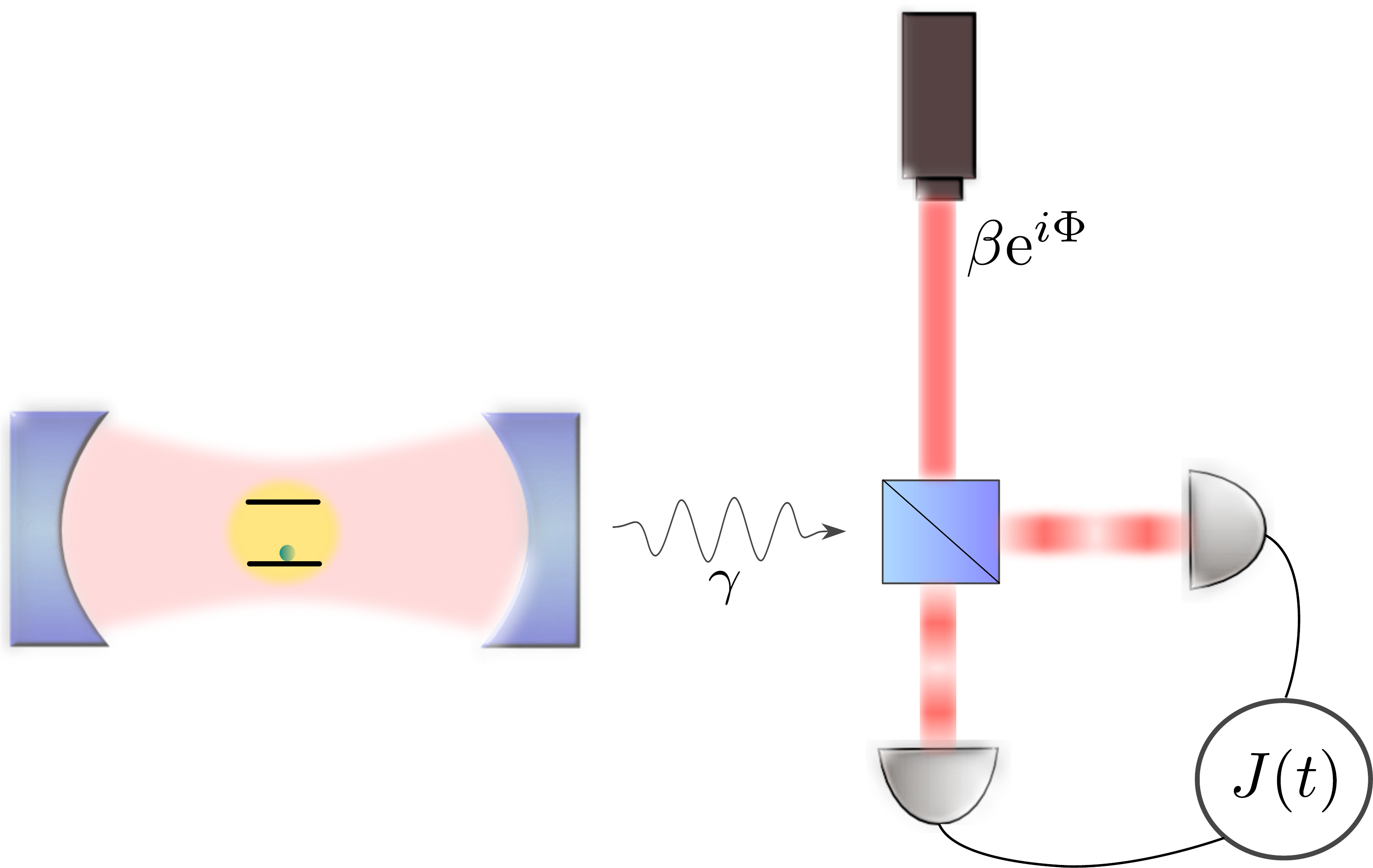}
}

\subfloat[\label{fig:current}]{
      \includegraphics[trim=0 0 0 0,width=0.95\columnwidth]{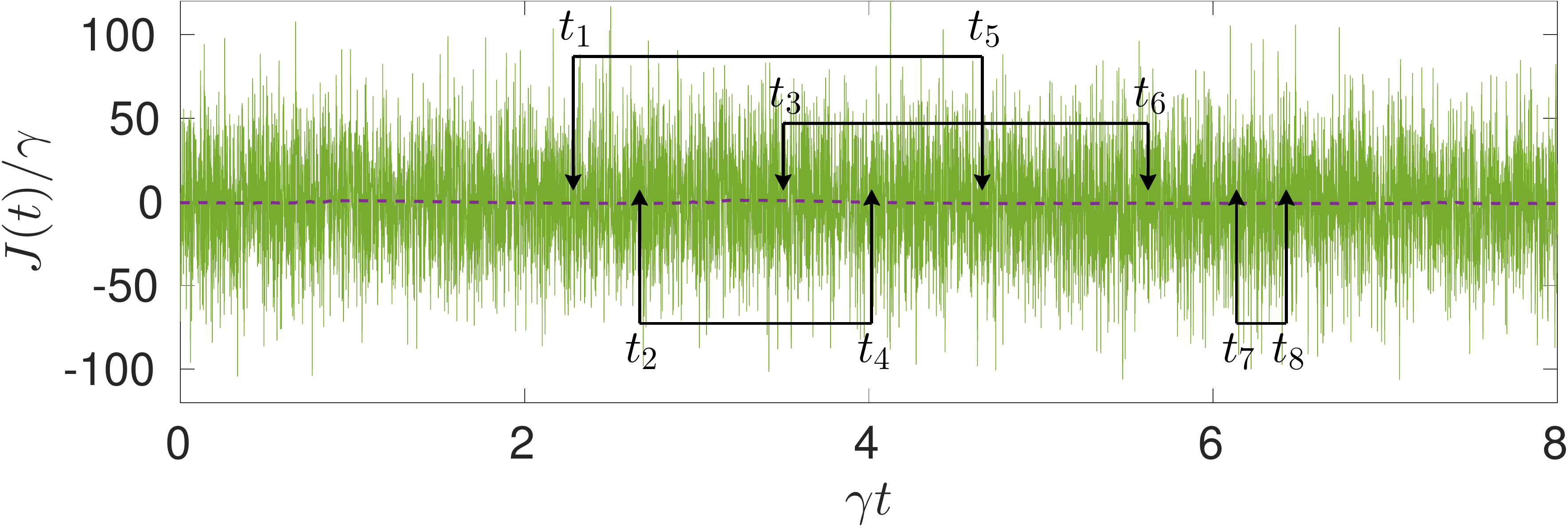}
    }

\subfloat[\label{fig:F1Mollow}]{
     \includegraphics[trim=0 0 0 0,width=0.95\columnwidth]{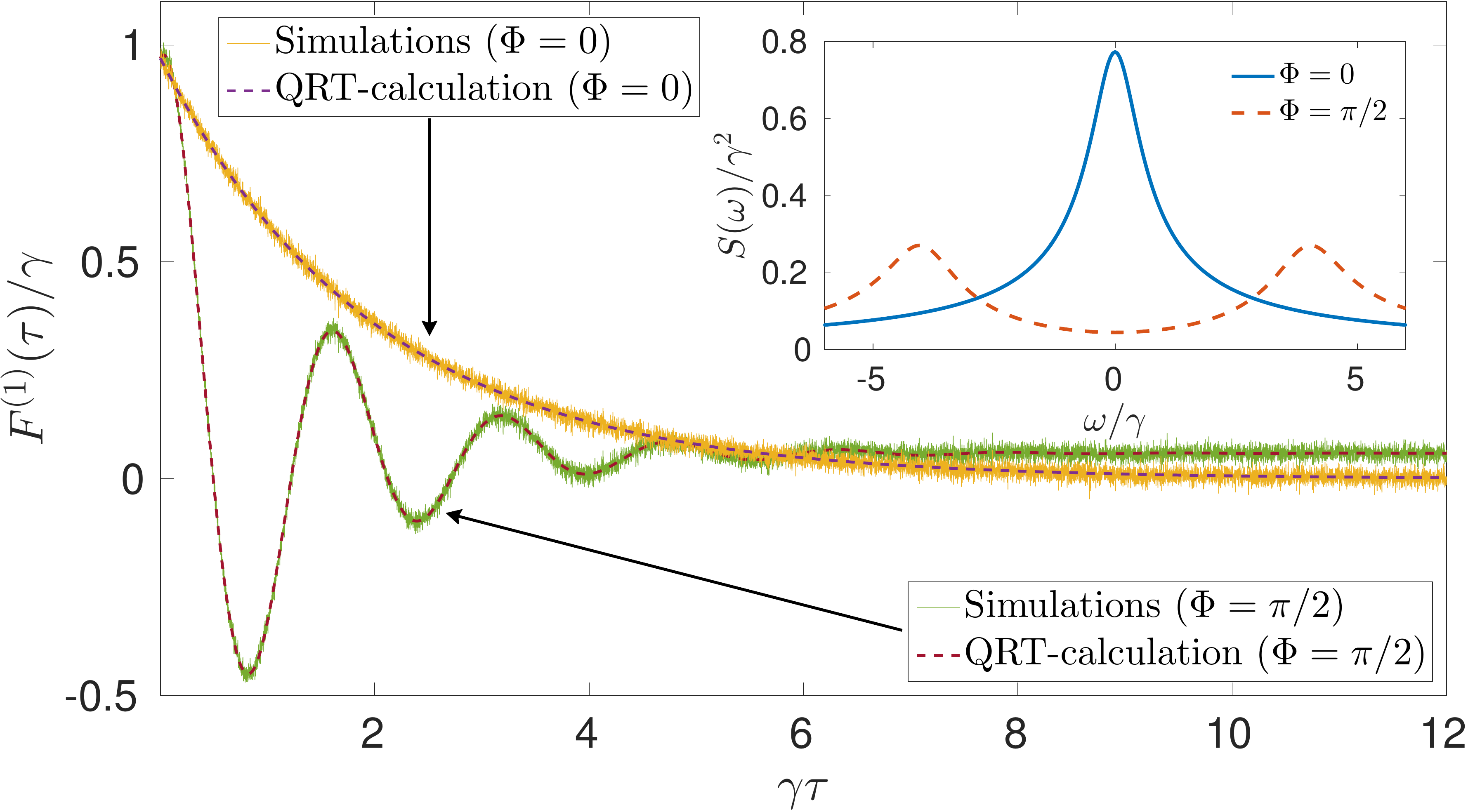}
    }
\caption{(Color online)
(a) Schematic experimental setup for balanced homodyne detection. The emission from the probed quantum system (here a two-level system in a cavity) is mixed with a strong local oscillator field with phase $\Phi$ in a 50/50 beamsplitter. The output ports are monitored by photo detectors and the homodyne current $J(t)$ is the difference between the two signals.
(b) An example of a homodyne current obtained by simulating the monitoring of a two-level system according to Eqs. (\ref{eq:J}), (\ref{eq:1}) is shown as the noisy green curve. The mean signal is indicated by the dashed purple line. The signal shows non-trivial temporal correlations, e.g., between the different times connected with arrows in the figure.
(c) The two-time homodyne current correlations \eqref{eq:Cl} averaged over 5000 independent realizations of $J(t)$. The quantum regression theorem \eqref{eq:F1} yields the theoretical results shown with purple and red, dashed curves. The inset shows the power spectrum of the homodyne current \eqref{eq:Sw} for a Rabi frequency of $\Omega_0=4\gamma$ for two choices of the local oscillator phase, $\Phi=0$ ($\sx$-probing) and $\Phi=\pi/2$ ($\sy$-probing).
}
\label{fig:mollow}
\end{figure}

In this work, we investigate the alternative situation of continuous homodyne detection of the field emitted by a quantum system, see \fref{fig:setup}. It is theoretically interesting to study the achievements of homodyne detection for precision measurements as the character of the signal and the measurement backaction is very different from that of photon counting. Such a study is further motivated by the extensive use of homodyne detection in optics where it often offers high efficiency and practical advantages over photon counting and in probing of microwave fields, e.g., in circuit QED \cite{PhysRevA.75.032329,PhysRevB.68.064509}, where photon counters are not available.

In Sec. II, we briefly introduce the experimental situation and the quantum trajectory analysis of the system dynamics subject to continuous homodyne field detection.
In Sec. III, we recall the Fisher information analysis and the Cramér-Rao sensitivity bound  associated with measurements with given outcome probabilities. We show how the quantum trajectory analysis and Bayes' rule allow extraction of maximum information from time dependent homodyne measurement records, and we evaluate the Fisher information associated with the homodyne detection scheme. In Sec. IV we study in detail the contributions to the Fisher information from the mean signal and from the two-time correlation function of the homodyne detection record, and discuss the results for the estimation of a Rabi drive strength applied to a two-level system. Sec. V provides a conclusion and an outlook.

\section{Quantum trajectories for homodyne detection}
The evolution of an unobserved, open quantum system is governed by a master equation $d\rho/dt = \L \rho$, where ($\hbar=1$)
\begin{align}\label{eq:liovillian}
\mathcal{L}\rho = -i[\hat{H},\rho]+\sum_k\left( \C_k\rho \Cd_k -\frac{1}{2} \left\{\Cd_k \C_k,\rho \right\}\right).
\end{align}
$\mathcal{L}$ is referred to as the Liouvillian  and the $\hat{c}_k$ operators represent relaxation processes.
The unobserved system relaxes to the steady state $\rst$, obeying $\L \rst=0$, from which average properties of the emitted florescence can be determined. If the system output is monitored, the evolution of the density matrix is affected by the measurement backaction. Homodyne detection performs an amplitude measurement of the emitted radiation by mixing it with a strong local oscillator (see \fref{fig:setup}). The homodyne current varies according to the current state $\rho(t)$ of the system and random, shot-noise, fluctuations,
\begin{align}\label{eq:J}
J(t) = \sqrt{\eta}\Tr{\xPhi\rho(t)} + \frac{d W_t}{dt},
\end{align}
where the measurement operator $\mathcal{X}_\Phi\rho = \C\e{-i\Phi} \rho +\rho\Cd\e{i\Phi}$ is given by the atomic dipole lowering operator  $\hat{c}$, $\Phi$ is the phase of the local oscillator,  and $\eta$ is the efficiency of the photo detectors. Detector shot-noise is modelled by the infinitesimal Wiener increment $d W_t$ with average properties  \cite{KJstochastic}
\begin{align}\label{eq:noise}
\E{dW_t} = 0 \quad \text{and} \quad dW_t dW_{\tp} = \delta(t-\tp)(dt)^2 .
\end{align}
That is, $dW_t$ is a normal distributed stochastic element with zero mean and variance $dt$.
The evolution of the system is, in turn, conditioned on the measurement results \cite{QMC},
\begin{align}\label{eq:1}
d\rho &= \L\rho dt+\sqrt{\eta}dW_t\left[\xPhi-\Tr{\xPhi\rho}\right]\rho,
\end{align}
where the average, deterministic evolution is governed by the Liouvillian \eqref{eq:liovillian} and the latter term accounts for the stochastic measurement backaction.

While our formalism and general analysis are valid for any quantum system, we shall exemplify the results and methods by considering estimation of the Rabi frequency of monochromatic laser driving of a single atomic system.
In a frame rotating with the driving frequency, the Hamiltonian is
\begin{align}\label{eq:H}
\hat{H}&=-\delta\sp\sm+\frac{\Omega}{2}\left(\sm+\sp \right),
\end{align}
where $\Omega$ is the Rabi frequency and $\delta$ is the laser-atom detuning.
If the system decays from the excited to the ground state at a rate $\gamma$, we have
$\hat{c} = \sqrt{\gamma}\sigma_{-}$. By adjusting the local oscillator phase $\Phi$, one may choose which spin component $\hat{\sigma}_\Phi = \cos \Phi \sx -\sin\Phi \sy$ is effectively probed. For $\Phi=0(\pi/2)$ in particular, the $\sx(\sy)$-component of the spin is measured. It follows from \eqref{eq:J} that the measurement backaction in \eqref{eq:1} corresponding to a particular phase $\Phi$ causes a rotation of the spin towards the axis defined by $\hat{\sigma}_\Phi $.

\section{Fisher information and Bayesian inference}
\label{sec:Bayes}
In quantum physics, measurement outcome data $D$, are governed by probabilities, assigned by the state of the system and being, hence, conditioned on the value of any unknown physical parameter $\theta$, governing the system dynamics. The more strongly $P(D|\theta)$ depends on $\theta$, the better is our ability to estimate that parameter. This is quantified by the Fisher information,
\begin{align}
\label{eq:fisher}
\mathcal{I}(\theta) &=
 \E{\left(\pdiff{\ln P(D|\theta)}{\theta}\right)^2 }.
\end{align}
Here $\E{\cdot}$ denotes the expectation value with respect to independent realizations of $D$. Note that while the underlying dynamics of the system and $P(D|\theta)$ may be governed by the laws of quantum physics, a theory dealing with the probability distribution of (classical) results of measurements is the same as for classical problems with stochastic measurement outcomes. Equation (\ref{eq:fisher}) is thus denoted the classical Fisher information which provides, via the (classical) Cramér Rao bound (CRB) \cite{Cramer}, a lower bound to the statistical variance $[\Delta S_\theta(D)]^2$ on any unbiased estimate $S_\theta(D)$ of $\theta$,
\begin{align}\label{eq:CRB}
[\Delta S_\theta(D)]^2\geq \frac{1}{\mathcal{I}(\theta)},
\end{align}
and yields the asymptotic precision of the best possible estimate.

The optimal estimate saturates the CRB \eqref{eq:CRB}, and is constructed by maximizing the probability that the unknown parameter has a given value $\theta$ conditioned on the outcome $D$, $S_\theta(D) = \text{max}_\theta\left[P(\theta|D)\right]$, \cite{Fisher}.
Bayes rule, in turn, defines an update rule for the prior (say uniform) probability $P(\theta)$,
\begin{align}\label{eq:Bayes}
P(\theta|D) = \frac{P(D|\theta)P(\theta)}{P(D)},
\end{align}
and hence delivers, by the probability factors assigned to the outcome $D$, an optimal parameter estimation strategy \cite{PhysRevA.87.032115}.

For multiple unknown variables, $\theta$ is a vector, and the Fisher information \textit{matrix} is $\mathcal{I}(\theta)_{nm} = \E{\pdiff{\ln P(D|\theta)}{\theta_n}\pdiff{\ln P(D|\theta)}{\theta_m}}$. The CRB then states a lower bound for the covariance in estimating pairs of parameters $\theta_n$ and $\theta_m$: $\mathrm{cov}\left[S(\theta_n),S(\theta_m)\right]\geq 1/\mathcal{I}_{nm}(\theta)$.
Generalization of the analysis in this paper to the multi-variable case is straightforward, but for notational purposes we shall restrict our attention to a single unknown parameter.

Since the classical Fisher information in \eqref{eq:fisher} refers to the probability distribution of the outcome of data that have been obtained by a \textit{specific} measurement scheme, it is clear that one may obtain more or less information by measuring different quantities, e.g., by employing photon counting rather than homodyne detection  or by choosing different values of the oscillator phase $\Phi$ in homodyne detection. These measurement schemes are mutually exclusive, and by obtaining information about one property, the experimentalist restricts himself or herself from measuring complementary observables of the emitted radiation field. An upper limit to the classical Fisher information is given by the quantum Fisher information (QFI), which quantifies the distinguishability of the state of the un-measured quantum system subject to different values of the sought parameter. For a system subject to a single measurement, this quantity is determined from the possible values of the system density matrix, while for our case of continuous probing of an emitted field, the distinguishability is governed by the entangled states of the system and the surrounding \textit{un-measured} quantized radiation field.
The ultimate ability to distinguish full quantum states of the emitter and the surrounding quantized field is given by the overlap of the states in question \cite{PhysRevLett.114.040401}, and the optimal measurement distinguishing such states may be difficult or even impossible to perform in practice. But, the QFI can be calculated by a simple master equation analysis \cite{PhysRevLett.112.170401,PhysRevA.93.022103}. For the radiatively damped two-level system, subject to resonant driving by an unknown Rabi frequency, we have carried out this calculation and found that the quantum Fisher information for determination of the Rabi frequency $\Omega$ is $4T/\gamma$, which is, notably, independent of the actual value of $\Omega$.

\subsection{Parameter estimation by Bayes rule}
\begin{figure}
\centering
\subfloat[$\Phi = \pi/2$\label{subfig:bayesY}]{
      \includegraphics[trim=0 0 0 0,width=0.9\columnwidth]{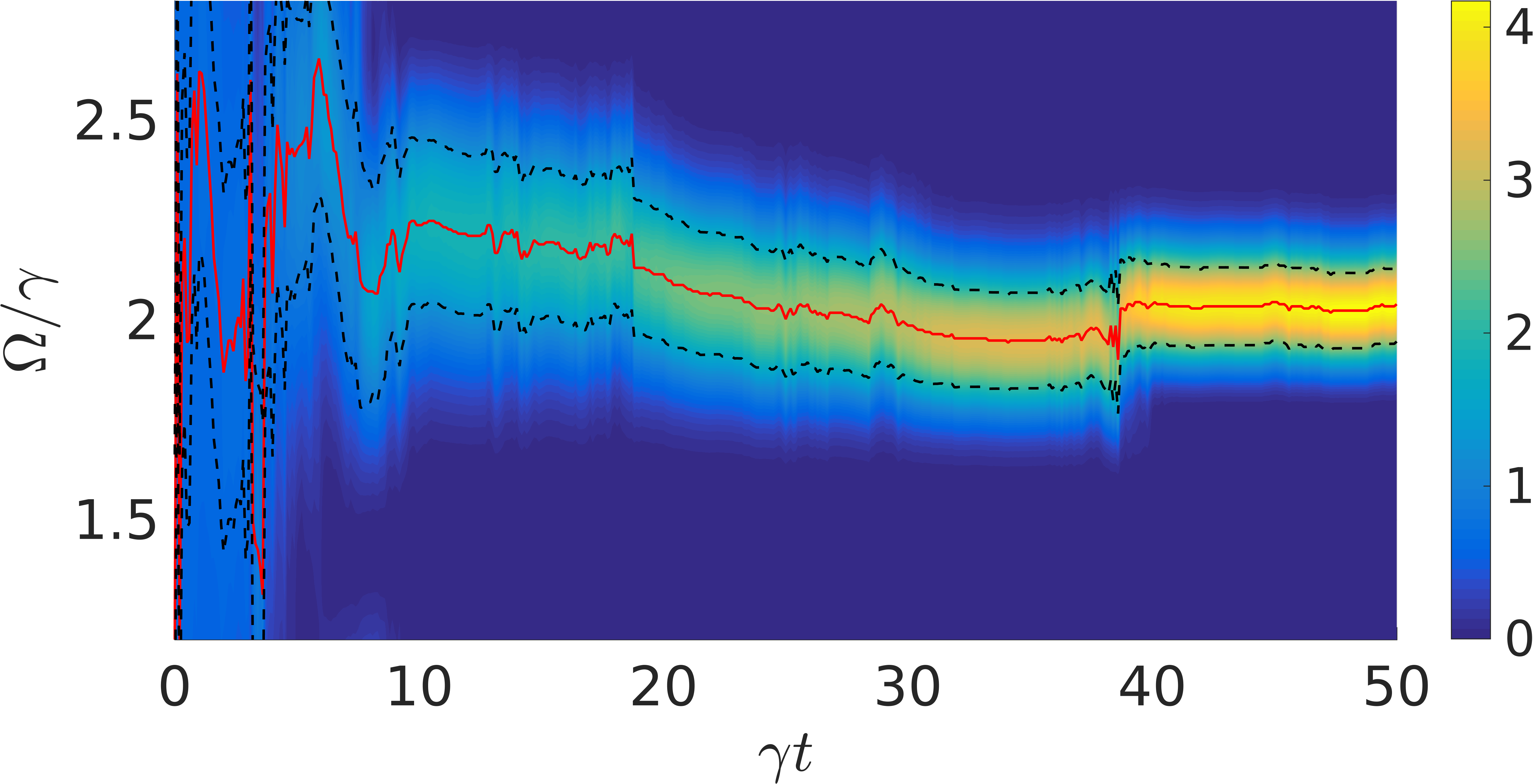}
    }
    \\
    \subfloat[$\Phi = 0$\label{subfig:bayesX}]{
      \includegraphics[trim=0 0 0 0,width=0.9\columnwidth]{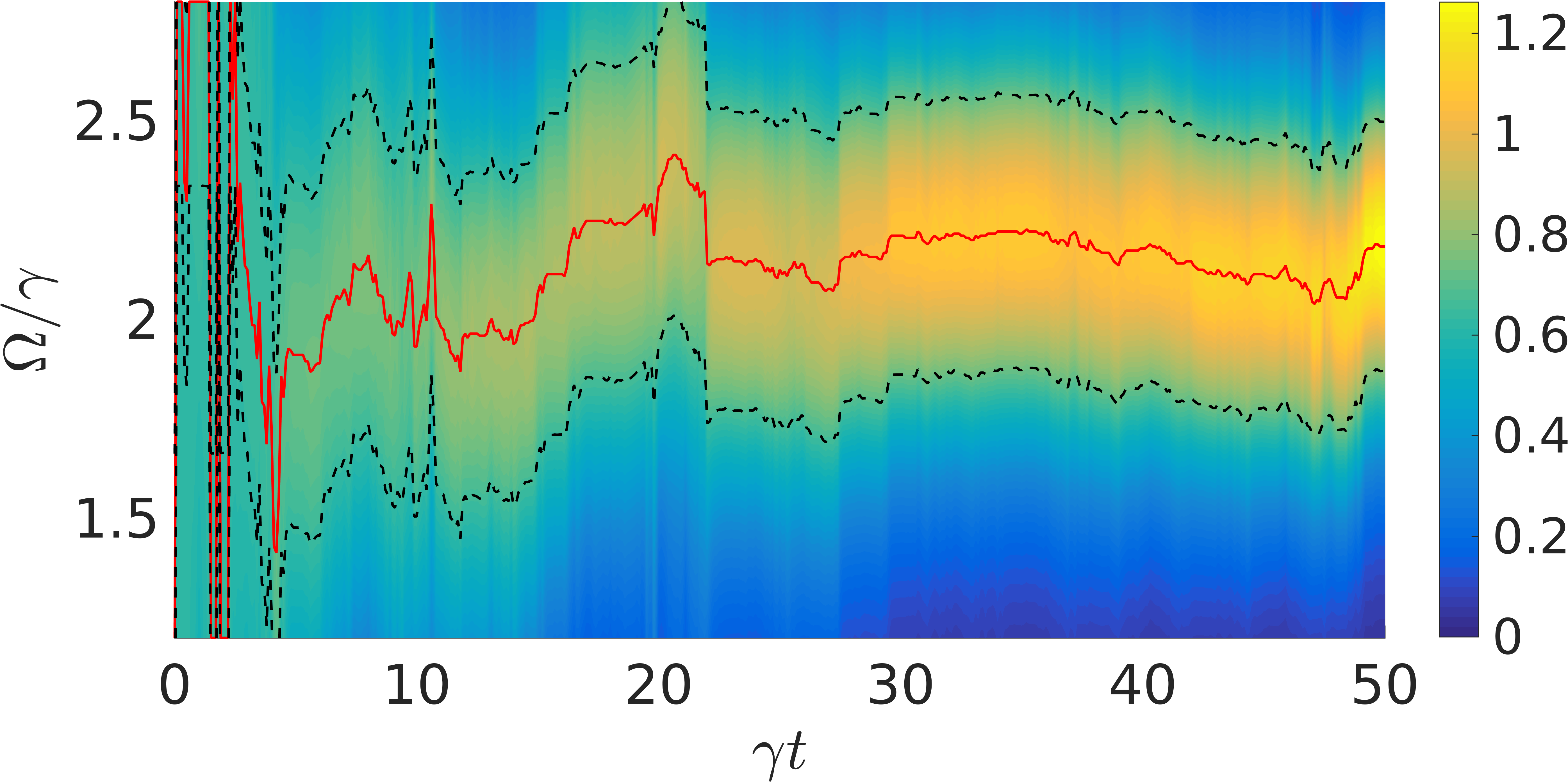}
    }
\caption{(Color online)
The evolution of the quasi continuous probability distribution $P(\Omega|D_t)$ for the Rabi frequency of a resonantly driven two-level system, conditioned on noisy homodyne measurement records [\fref{fig:current}].
The red lines track the most likely parameter value $S_\Omega(D_t) = \mathrm{max}_\Omega\left[ P(\Omega|D_t)\right]$, and the dashed, black lines show the FWHM of $P(\Omega|D_t)$.
Results are shown for two choices of the local oscillator phase, (a) $\Phi=\pi/2$ ($\sy$-probing) and (b) $\Phi=0$, ($\sx$-probing). The true value $\Omega_0=2\gamma$, assumed for the Rabi frequency, is gradually identified in each simulation.
}
\label{fig:bayes}
\end{figure}

In a continuous homodyne measurement, the output data is the stochastic measurement current, $D_t = \{J(\tp)|0\leq \tp \leq t\}$. The instantaneous signal value \eqref{eq:J} is dominated by white noise, see \fref{fig:current}, which affects the  subsequent evolution \eqref{eq:1}. The probability of the measurement signal at each instant can be calculated by following the quantum trajectory and determining at each time step the probability for the actually measured (in our study: simulated) current. The probability for the whole measurement record, accumulated this way, can conveniently be determined by solving instead the linear master equation
\begin{align}\label{eq:rho_bar}
\frac{d\bar{\rho}}{dt} = \mathcal{L}\bar{\rho} +\sqrt{\eta}J(t)\xPhi\bar{\rho}
\end{align}
for an un-normalized density matrix $\bar{\rho}$: $P(J(t)|\theta)\propto \Tr{\bar{\rho}}$.
Propagation of $\bar{\rho}$ for all candidates $\theta$, but conditioned on the actual experimental outcome $J(t)$, thus provides via \eqref{eq:Bayes} an update rule for the probability distribution $P(\theta|D_t)$ \cite{PhysRevA.87.032115}.

In \fref{fig:bayes}, we have simulated this procedure.
The red lines track the best estimate $S_\theta(D_t) = \mathrm{max}_\theta\left[P(\theta|D_t)\right]$, while the dashed, black lines indicate  the width of the probability distribution.
For probing of $\sy$ in \fref{subfig:bayesY}, the distribution quickly becomes normal as demanded by the central limit theorem, and we observe a smooth convergence of the most likely value around the true $\Omega_0=2\gamma$ used to simulate the measurement record $J(t)$. For probing of $\sx$, shown in panel \fref{subfig:bayesX}, the convergence is much slower, and as reflected by the broader distribution at $\gamma t=50$, less information is gained per time.
This is due to the fact that on resonance ($\delta=0$) the Hamiltonian \eqref{eq:H} commutes with the $\sx$-operator, and as we shall discuss below, the finite information stems from higher order temporal correlations in the signal decaying with a $\Omega$-dependent rate.


The uncertainties in the estimates are provided by the CRB \eqref{eq:CRB} with the Fisher information of the full signal given by

\begin{align}\label{eq:FI}
\mathcal{I}(\theta) = \mathbb{E}\left[\frac{1}{\Tr{\bar{\rho}}^2}\left(\frac{\partial \Tr{\bar{\rho}}}{\partial\theta}\right)^2\right].
\end{align}
By defining
$
\zeta = \frac{1}{\Tr{\bar{\rho}}}\frac{\partial \bar{\rho}}{\partial\theta_i},
$
the Fisher score is $\Tr{\zeta}$ \cite{PatternClassification}, and the equation of evolution derived from \eqref{eq:rho_bar},
\begin{align}\label{eq:dZeta}
\begin{split}
d \zeta &= \left(\mathcal{L}\zeta +\pdiff{\mathcal{L}}{\theta}\rho\right) dt
\\
&+\sqrt{\eta}dW_t\left(\xPhi \zeta+\pdiff{\xPhi}{\theta}\rho-\Tr{\xPhi\rho}\zeta\right),
\end{split}
\end{align}
allows calculation of the classical Fisher information,
$\mathcal{I}(\theta)=\E{\Tr{\zeta}^2}$ without numerical evaluation of derivatives of noisy quantities \cite{PhysRevA.87.032115}.
Note that we must simultaneously solve \eqref{eq:1} for the accompanying normalized state $\rho(t)$. Unlike operator expectation values for which the average over trajectories is equivalent to results obtained by the unconditioned density matrix, we do not have a deterministic theory for the average of a nonlinear expression such as $\E{\Tr{\zeta}^2}$. Equation (\ref{eq:dZeta}) may, however, be simulated in a trajectory analysis and the Fisher information \eqref{eq:FI} obtained by averaging over many trajectories.
\begin{figure*}
\subfloat[\label{fig:Iphi}]{
\includegraphics[trim=0 0 0 0,width=0.65\columnwidth]{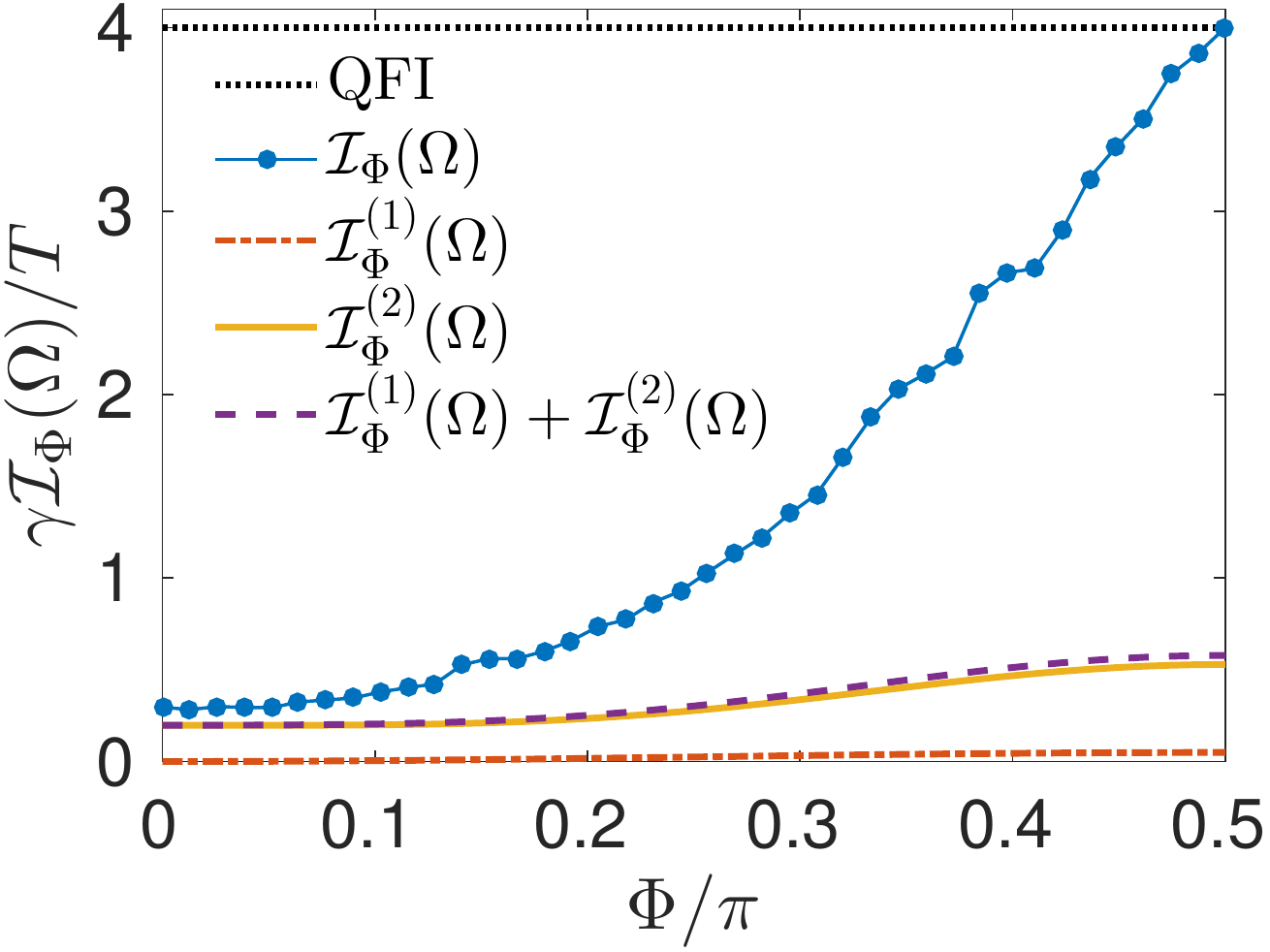}
}
\subfloat[\label{subfig:I}]{
\includegraphics[trim=0 0 0 0,width=0.65\columnwidth]{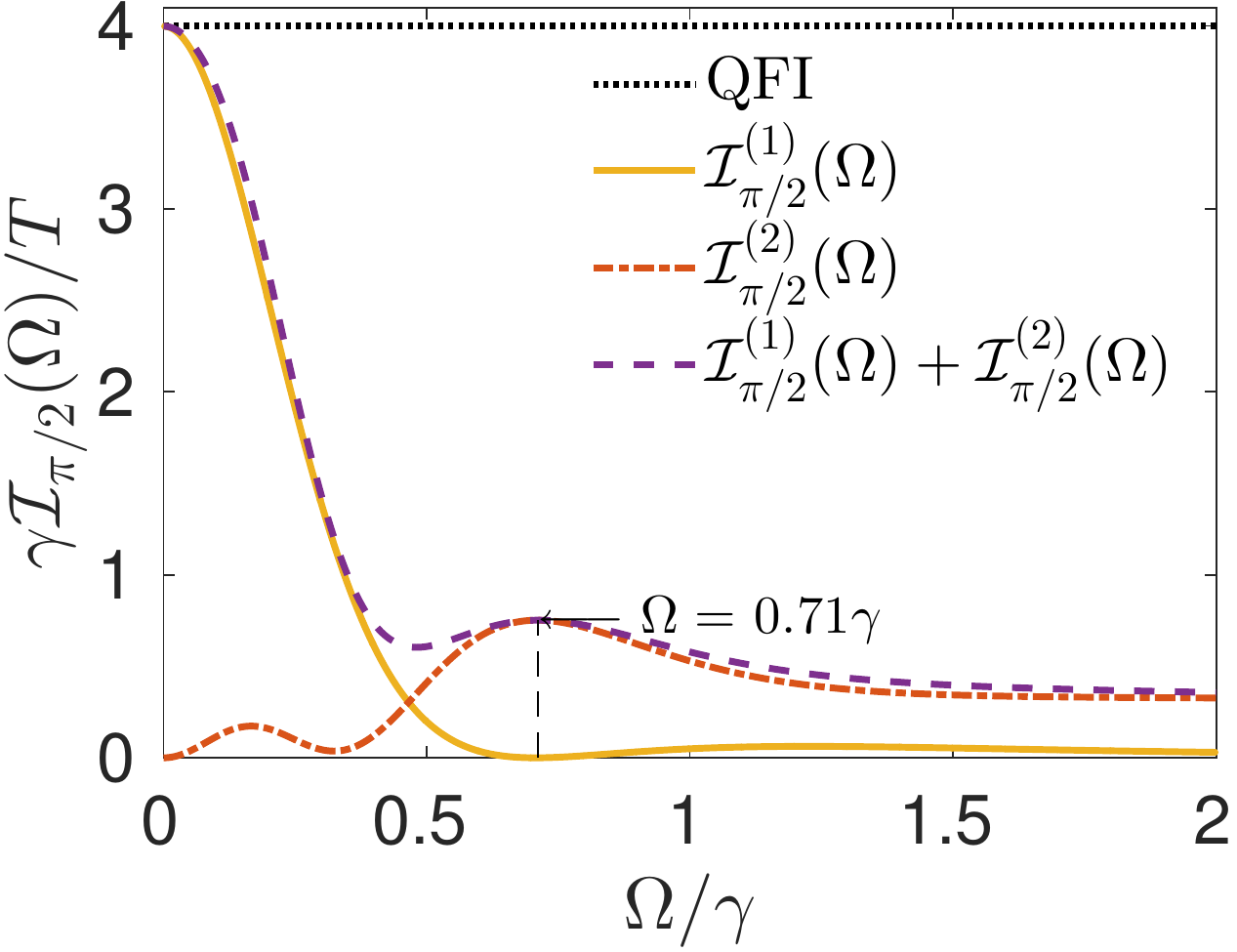}
}
\subfloat[\label{subfig:I2PhiOmega}]{
\includegraphics[trim=0 0 0 0,width=0.65\columnwidth]{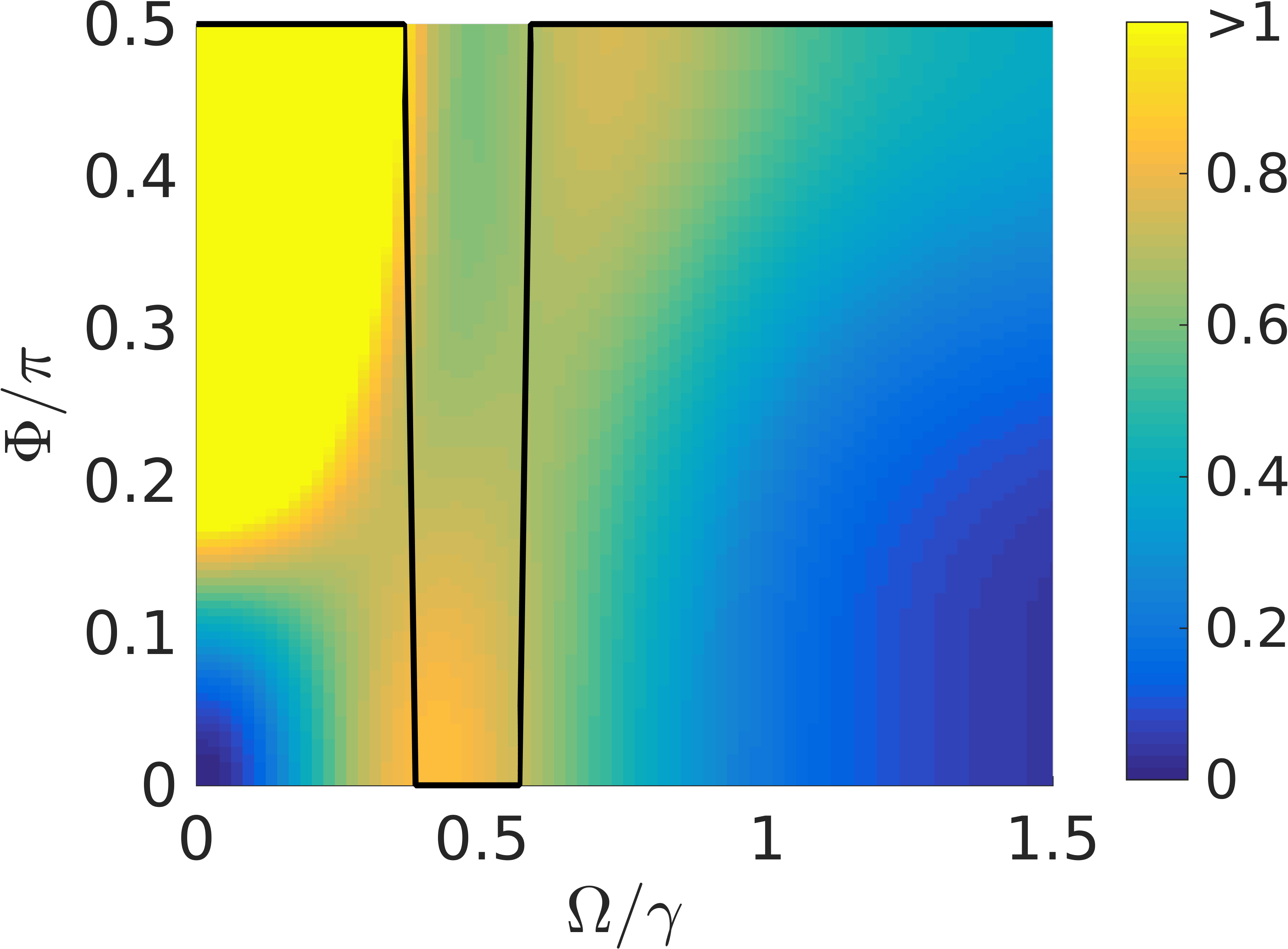}
}

\caption{(Color online)
Contributions to the Fisher information $\mathcal{I}_\Phi(\Omega)$ per time $T$ for Rabi frequency estimation in a resonantly driven two-level system. The Fisher information is upper bounded by the value $4T/\gamma$ of the QFI, which is independent of the Rabi frequency, and it is lower bounded by the information retrieved by the integrated (mean) signal $\Iet_\Phi(\Omega)$.
$\Ito_\Phi(\Omega)$ describes the information provided by the two-time correlations in the homodyne current. Results are shown in
(a) as a function of the local oscillator phase $\Phi$ for $\Omega=\gamma$ and in (b) as a function of the Rabi frequency $\Omega$ for $\Phi=\pi/2$.
In (c), the value of $\Iet_\Phi(\Omega)+\Ito_\Phi(\Omega)$ per time $T$ is indicated by the color scheme as a function of $\Phi$ and $\Omega$, and the maximum value is tracked by the black line.
}
\end{figure*}
In \fref{fig:Iphi}, we show an example of such a calculation with $100\,000$ trajectories for the Fisher information $\mathcal{I}(\Omega)$ as a function of the local oscillator phase. We find that asymptotically, the Fisher information grows linearly with time $T$, and the figure shows the value of the scaled quantity $\gamma \mathcal{I}(\Omega)/T$. As anticipated from the Bayesian studies in \fref{fig:bayes}, the Fisher information takes its largest values when $\Phi\simeq \pi/2$, where it even reaches the quantum Fisher information. Our numerical simulations of the two level system dynamics for $\Phi=\pi/2$, show that, for all values of the Rabi frequency $\Omega$, homodyne detection indeed reaches a Fisher information of $4T/\gamma$, and is identical to the QFI. This result is illustrated in \fref{fig:FisherEta}.

To gain analytical insight in the results, we consider the case of weak driving ($\Omega\ll \gamma$), where we can use the Holstein-Primakoff approximation \cite{PhysRev.58.1098} and replace the two level system by a driven oscillator [$\sm\rightarrow \hat{a}$ in \eqref{eq:H}, where $\hat{a}$ is a bosonic annihilation operator]. Thanks to the simple algebra of oscillator observables, \eqref{eq:rho_bar} may be solved exactly \cite{Unpublished},
\begin{align}
\begin{split}
\bar{\rho}(t) &=
\e{
(\mathcal{L}-\frac{1}{2}\xPhi^2)t
}
\e{
\Int{0}{t}{t}{J(t^\prime)\tilde{\mathcal{X}}_\Phi(t^\prime)}
}
\\
&\times
\e{
\frac{\Omega \sqrt{\gamma}}{\delta^2+(\gamma/2)^2} \int_0^t dt'\, J(t)\e{-\gamma/2 t'}\left[ \delta\cos(\delta t'+\Phi)+\frac{\gamma}{2}\sin(\delta t'+\Phi)\right]
}
\bar{\rho}(0),
\end{split}
\end{align}
where
$
\tilde{\mathcal{X}}_\Phi(t)\rho = \left(\e{-i(\delta t+\Phi)}\hat{c}\rho+
\rho\hat{c}^\dagger\e{i(\delta t+\Phi)}\right)\e{-\frac{\gamma}{2}t}.
$
Picking the initial state $\bar{\rho}(t=0)$ as the coherent steady state $\ket{-i\frac{\Omega}{\gamma}}$, the Fisher information \eqref{eq:FI} is readily obtained and it yields the result $\mathcal{I}(\Omega) = 4T/\gamma$, as expected.



\section{Properties of the homodyne signal}
It is remarkable that at resonant driving ($\delta=0$) the Bayesian analysis of the homodyne detection signal has the same sensitivity for all values of the Rabi frequency, and that it completely exhausts the information about the Rabi frequency present in the quantized multi-mode radiation field.
As stated in Sec. I, this ability springs from the fact that we are not only using the integrated signal but also the temporal correlations established by the  measurement backaction during the continuous probing of the system.
We refer the reader to Ref. \cite{PhysRevLett.112.170401} for an analysis of the case of finite laser-atom detuning.


The instantaneous homodyne current is dominated by noise, and it provides useful information only when it is integrated over a finite time. We define the average signal from the time $t=0$ until time $t=T$,
\begin{align}\label{eq:I}
Y\equiv \lim_{T\rightarrow \infty}\frac{1}{T}\Int{0}{T}{t}{J(t)},
\end{align}
with the mean value
\begin{align}\label{eq:Ecurrent}
I = \E{Y} = \sqrt{\eta}\Tr{\xPhi\rst}.
\end{align}
Note that due to \eqref{eq:J}, $J(t)-\E{J(t)}= \frac{dW_t}{dt}$.
Hence, the variance on the integrated current stems from white noise, \eqref{eq:noise}, and we find
\begin{align}\label{eq:varI}
\mathrm{var}(Y) = \frac{1}{T}.
\end{align}
The integrated signal thus allows extraction of the average system properties with a relative  error decreasing as $1/\sqrt{T}$, and the same temporal scaling applies to the estimation error for any system parameter. Due to saturation, however, the steady state properties of the signal may depend only very weakly on the Rabi driving frequency. This is where correlations within the signal may hold further important information due to the transients triggered by the backaction of the continuous homodyne measurement. We shall here specifically address the two-time correlation function of the homodyne current to study the information gained by including two-time correlations together with the integrated signal in the analysis.
While we choose to consider this restriction for matter of analysis, we note that experimental hardware may provide a data record that has been reduced to provide only the two-time correlation function, equivalent to the power spectrum of the homodyne current which can be  directly sampled by a spectrum analyzer.
%

\subsection{Two-time correlations}
Two-time correlations in the signal are extracted as the average value
\begin{align}\label{eq:Cl}
\Cl \equiv \lim_{T\rightarrow \infty}\frac{1}{T-\tau}\Int{0}{T-\tau}{t}{ J(t+\tau)J(t)},
\end{align}
where, in the limit of long probing times, the initial state may be taken as the steady state, and $\Cl$ depends only on the time-difference $\tau$.

Repeating the experiment (or repeating the quantum trajectory simulation) yields $N$ independent samples of the current $J(t)$
and thereby $N$ independent realizations of the integrated signal \eqref{eq:I} and the autocorrelation function \eqref{eq:Cl}.
In \fref{fig:F1Mollow}, the two-time correlations \eqref{eq:Cl} are averaged over $5000$ such simulated homodyne currents. Note the oscillatory behavior for $\sy$-probing while $\sx$-probing merely triggers an exponentially decaying transient.

The average over many independent samples of the homodyne current $J(t)$ can be calculated deterministically by the quantum regression theorem (QRT),
\begin{align}\label{eq:F1}
F^{(1)}(\tau) \equiv \mathbb{E}\left[C(\tau)\right] &= \eta\Tr{\xPhi\eLt{\tau}\xPhi\rst}+\delta(\tau),
\end{align}
and it perfectly matches  the simulated data in \fref{fig:F1Mollow}.

The Fourier transform of the two-time correlation function is the power spectrum $\Sw$ of the homodyne signal,
\begin{align}\label{eq:Sw}
S(\omega) = \Int{-\infty}{\infty}{\tau}{F^{(1)}(\tau) \e{-i\omega \tau}}.
\end{align}
As seen in the inset of \fref{fig:F1Mollow}, $\sx$-probing yields a single frequency peak at the atomic resonance frequency, while $\sy$-probing gives a signal with frequency components at $\pm \Omega$ and hence yields much more information on the Rabi frequency.
The widths of the peaks and equivalently the decay rates of the two-time correlation functions in \fref{fig:F1Mollow}, however, depend weakly on the Rabi frequency. This  is what allows information to be obtained also by $\sx$-probing and constitutes the basis for the slow convergence of the Bayesian estimate in \fref{subfig:bayesX}.

Together the spectra reflect the three peaks of the well-known Mollow triplet of florescence from a driven two-level emitter \cite{mollow}.

\subsection{Fisher information from two-time correlations}
We shall now evaluate the Fisher information associated with the integrated signal and the two-time correlations in the homodyne signal current, as if only those quantities are made available for the estimation of the unknown parameter $\theta$. In a recent article \cite{Burgarth2015} that considers parameter estimation by discrete, sequential measurements on a quantum system, it is shown that as the number $N$ of measurement data increases, the probability distribution of functionals of the data (e.g., mean values and  multi-time correlations) become asymptotically normal.

Homodyne detection is a continuous measurement, and to apply the results of Ref. \cite{Burgarth2015} to our problem,  we consider initially the current $J(t_i)$ at $N$ discrete times $t_i$, where $t_{i+1}-t_i=\Delta t=T/N$.
Then
\begin{align}
\begin{split}
Y &= \frac{1}{N}\sum_{i=1}^N J(t_i),
\\ 
C_l &= \frac{1}{N-l}\sum_{i=1}^{N-l} J(t_i)J(t_i+t_l),
\end{split}
\end{align}
approximate Eqs. (\ref{eq:I},\ref{eq:Cl}), and
\cite{Burgarth2015} assures that $\vec{X} = (I,C_1,...,C_{L})^T$ is a multivariate Gaussian random variable, i.e., the probability $P(\vec{X})$ becomes asymptotically normal for large $N$,
\begin{align}\label{eq:prob}
P(\vec{X}) \propto \e{\frac{1}{2}\left(\vec{X}-\E{\vec{X}}\right)^\mathrm{T}\vec{\Sigma}^{-1}\left(\vec{X}-\E{\vec{X}}\right)}.
\end{align}
Here the covariance matrix of $\vec{X}$ is defined as
\begin{align}
\Sigma_{ij} &=\E{\left(X_i-\E{X_i}\right) \left(X_j-\E{X_j}\right)}.
\end{align}
\begin{widetext}
The first diagonal element of the covariance matrix is given simply by \eqref{eq:varI}, $\Sigma_{00}=1/T$, while the evaluation of the covariances between the sampled autocorrelations with different time delays $\tau$ involves up to four-time correlations in the measurement signal,
\begin{align}\label{eq:Sigmallp}
\Sigma(\tau,\tau^\prime) = \lim_{T\rightarrow \infty}\frac{1}{(T-\tau)^2}\int_0^{T-\tau} dt \int_0^{T-\tau} dt^\prime \, \E{\left(J(t+\tau)J(t)-\Fet(\tau)\right)\left(J(\tp+\tau^\prime)J(\tp)-\Fet(\tau^\prime)\right)}.\end{align}
The four-time correlations, and more general multi-time correlation functions, are quantified by products of the current readout at multiple times with definite separation intervals, and when averaged over the time argument of the first readout and over many independent realizations, they read
\begin{align}\label{eq:multi}
F_{\Phi}^{(n/2)}(\{\tau_j\})&\equiv \braket{\prod _{i=1}^n J(\sum_{1\leq j\leq i} \tau_j)}
=
\Tr{\prod _{i=1}^n\left( \xPhi\e{\mathcal{L}\tau_i}\right)\xPhi\rho_{\text{st}}},
\qquad\text{for }\quad \tau_i>0,
\end{align}
where $\braket{\cdot}$ denotes an average over the time $\tau_1 $.
The covariance matrix elements thus follow from time-ordered integration of four-time correlation functions given in \eqref{eq:multi}. Notice, however, that this seemingly complicated task becomes remarkably simple due to the noise properties of the homodyne signal, see \fref{fig:current}. In fact,
the leading contribution to the integrand stems from the noise terms in \eqref{eq:J} when the time arguments match two by two in \eqref{eq:Sigmallp}, i.e. to leading order in $1/dt$ we obtain,
\begin{align}\label{eq:covar}
\Sigma(\tau,\tau^\prime)&\simeq  \lim_{T\rightarrow \infty}\frac{1}{(T-\tau)^2}\int_0^{T-\tau} dt \int_0^{T-\tau} dt^\prime \, \mathbb{E}\left[\frac{dW(t)}{dt}\frac{dW(t+\tau)}{dt}\frac{dW(\tp)}{dt}\frac{dW(\tp+\tau^\prime)}{dt}\right]
\nonumber\\
&=\lim_{T\rightarrow \infty}\frac{1}{(T-\tau)^2}\int_0^{T-\tau} dt \int_0^{T-\tau} dt^\prime \,
\delta(t-\tp)\delta(\tau-\tau^\prime)
=
\frac{\delta(\tau-\tau^\prime)}{T},
\end{align}
where we applied \eqref{eq:noise} at the second step.
\end{widetext}
Equation (\ref{eq:covar}) implies
$
\Sigma_{ij} = \frac{\delta_{ij}}{T \Delta t} 
$
for $i,j>0$.
Remarkably, the covariance matrix is diagonal, and to leading order in $1/T$ there is no covariance between the autocorrelations at different times. We have verified this numerically by the covariances between the simulated data shown in \fref{fig:F1Mollow}.
The variance increases as the numerical $C_l$-grid is made finer, leading to a convergence of the information as $\Delta t$ is decreased.

As seen in \fref{fig:F1Mollow} the two-time correlation function \eqref{eq:F1} approaches $I^2$ for large correlation times $\tau$ and suggests that the mean current and the asymptotic correlation are correlated. To avoid the trivial covariance between $Y$ and $C(\tau)$ we therefore subtract the expected integrated signal contribution from the correlations,
\begin{align}
\Cl\rightarrow \Cl-I^2 \quad \text{and} \quad \Fet(\tau)\rightarrow \Fet(\tau)-I^2.
\end{align}
Without omitting any contributions to the total information acquired from the experiments, this ensures that $\Sigma_{0l}=\Sigma_{l0} = 0$.


The Fisher information matrix of a multivariate normal distribution is well-known,
\begin{align}\label{eq:FisherMultivariate}
\mathcal{I}_{\text{MV}}(\theta) =
\pdiff{\E{\vec{X}}^\mathrm{T}}{\theta}
\vec{\Sigma}^{-1}
\pdiff{\E{\vec{X}}^\mathrm{T}}{\theta}
+\frac{1}{2}\Tr{\vec{\Sigma}^{-1}\pdiff{\vec{\Sigma}}{\theta}\vec{\Sigma}^{-1}\pdiff{\vec{\Sigma}}{\theta}}.
\end{align}
The inverted covariance matrix is proportional to the total time of probing,
and since the latter term in \eqref{eq:FisherMultivariate} does not scale with $T$ and hence becomes asymptotically negligible, we find
$
\mathcal{I}_{\text{MV}}(\theta) = \mathcal{I}^{(1)}(\theta) + \mathcal{I}^{(2)}(\theta),
$
where
\begin{align}
\label{eq:I1}
\mathcal{I}^{(1)}(\theta) &= T\left(\pdiff{I}{\theta}\right)^2
\\
\mathcal{I}^{(2)}(\theta)
&= T\sum_i \left(\frac{\partial \mathbf{F}}{\partial \theta}\right)_i^2 \Delta \tau.
\end{align}
Equation (\ref{eq:I1}) yields the contribution to the Fisher information from the integrated signal, and we may finally take the continuum limit $\Delta \tau \rightarrow d\tau$ to find the contribution from two-time signal correlations,
\begin{align}\label{eq:I2}
\mathcal{I}^{(2)}(\theta) =  T\int_{\tau_{\text{min}}}^{\tau_{\text{max}}} d\tau \left(\frac{\partial F^{(1)}(\tau)}{\partial \theta}\right)^2.
\end{align}
By
\eqref{eq:Sw}, a spectral analysis yields the same information as a direct analysis of the two-time correlated time series. In fact,
we may apply Plancherels theorem (unitarity of the Fourier transform) in \eqref{eq:I2} to obtain the expected Fisher information from the spectrum,
\begin{align}\label{eq:FisherW}
\mathcal{I}^{(2)}(\theta) =  T\int_{-\infty}^{\infty} d\omega \, \left(\frac{\partial S(\omega)}{\partial \theta}\right)^2.
\end{align}

\subsection{Achieving the Cramér-Rao Bound with a linear filter}
As proven by Fisher \cite{Fisher} a Bayesian analysis saturates the Cramér-Rao bound and allows parameter estimation with a precision given by the Fisher information.
The CRB may, however, equivalently be saturated by applying a linear filter to the data available for the estimation process.

The Fisher information associated with the integrated signal $Y$ and two-time correlations $C(\tau)$ takes a simple form Eqs. (\ref{eq:I1},\ref{eq:I2}), and for a given experimental realization of $Y$ and $C(\tau)$, an unbiased estimator $S_\theta(D)$ for which the variance is minimized and the CRB reached exists. Suppose that asymptotically the parameter value has been identified to a small vicinity $\delta\theta$ around a value $\theta_0$. With the Fisher information $\Iet(\theta_0)+\Ito(\theta_0)$ this estimator is given explicitly by the linear filter,
\begin{align}
\begin{split}
S_{\delta \theta}(D_T) = \frac{T}{\mathcal{I}^{(1)}(\theta_0)+\mathcal{I}^{(2)}(\theta_0)}\bigg[
\left. \frac{\partial I(\theta)}{\partial \theta}\right|_{\theta = \theta_0}\left(Y-I(\theta_0) \right)
\\
+
\int d\tau \,
\left.\frac{F^{(1)}(\tau,\theta)}{\partial \theta}\right|_{\theta = \theta_0}\left(C(\tau)-F^{(1)}(\tau,\theta_0)\right)
\bigg].
\end{split}
\end{align}
This expression represents a first order correction to the initial estimate $\theta_0$ according to the dissimilarity between the expected and the recorded signal, and by normalizing the estimate by the Fisher information per time, it ensures that larger uncertainties allow larger adjustments. The linear filter is valid in the asymptotic limit where $\delta\theta$ is small.

\subsection{Rabi frequency estimation}
We turn now to the example of Rabi frequency estimation in a resonantly driven two-level system [\eqref{eq:H} with $\delta=0$] where all calculations are straightforward. We consider first perfect detection ($\eta=1$).

The Fisher information from the mean signal \eqref{eq:I1} is
\begin{align}\label{eq:I1ex}
\frac{\mathcal{I}_\Phi^{(1)}(\Omega)}{T} =
4\gamma^3\left[\frac{2\Omega^2-\gamma^2}{(2\Omega^2+\gamma^2)}\right]\sin^2{\Phi}.
\end{align}
For $\Phi=0$, $\Fet(\tau)$ may be evaluated analytically,
\begin{align}\label{eq:F0}
F_0^{(1)}(\tau)  = \frac{2\Omega^2 \gamma \e{-\gamma\tau/2}}{2\Omega^2+\gamma^2},
\end{align}
leading to a Fisher information,
\begin{align}
\frac{\mathcal{I}_0^{(2)}(\Omega)}{T} = \frac{16\Omega^2 \gamma^6 }{(2\Omega^2+\gamma^2)^4}.
\end{align}
The general expression for $F^{(1)}_{\pi/2}$ is more complicated.
An example is shown in \fref{fig:F1Mollow}.

In \fref{fig:Iphi}, we compare $\Iet_\Phi(\Omega)$, $\Ito_\Phi(\Omega)$ and the total contribution for at most two-time correlations in the signal $\Iet_\Phi(\Omega)+\Ito_\Phi(\Omega)$ to the QFI and the Fisher information of the full signal \eqref{eq:FI}. Results are shown as a function of $\Phi$ for $\Omega=\gamma$.
Because the steady state solution in \eqref{eq:Ecurrent}
has no $\sx$-component, the information $\mathcal{I}_{\Phi}^{(1)}$ from the integrated signal is maximized for $\Phi=\pi/2$, while it vanishes for $\Phi=0$. As discussed after \eqref{eq:Sw} two-time correlations, on the other hand, are able to extract information from $\sx$-probing.

The information in the full signal $\mathcal{I}_\Phi(\Omega)$  in \fref{fig:Iphi} is much higher than the contributions from the integrated signal $\Iet_\Phi(\Omega)$ and two-time correlations $\Ito_\Phi(\Omega)$, reflecting that higher order correlations in the signal are responsible for the bulk of the information extracted by the Bayesian protocol.
For $\Phi=0$, however, the main part of the information stems from just two-time correlations.

In \fref{subfig:I2PhiOmega}, we show in colors the dependence of the Fisher information $\Iet_\Phi(\Omega)+\Ito_\Phi(\Omega)$ on the phase $\Phi$ of the local oscillator and on the applied Rabi frequency $\Omega$.
While the information from two-time correlations is maximal close to $\Omega\simeq 0.5\gamma$ for all $\Phi$, the integrated signal favours $\Omega = 0$ and $\Phi=0$. This leads to a combined Fisher information with three distinct maxima, and we track by the black line, the optimal local oscillator phase for different values of the Rabi frequency. Surprisingly, in the vicinity of $\Omega=0.5\gamma$, $\Phi=0$ is the optimal choice while in general $\Phi=\pi/2$ is optimal. There is no intermediate range, where an intermediate value of the phase $0<\Phi<\pi/2$ is favourable.

In \fref{subfig:I} the comparison is performed as a function of the actual value of the Rabi frequency for $\Phi = \pi/2$.
The information is upper bounded by the QFI and, as evidenced by the large discrepancy between the QFI and $\Iet_{\pi/2}(\Omega)+\Ito_{\pi/2}(\Omega)$, the main part of the full information comes from higher order correlations.

Below saturation ($\Omega< \gamma/2$) the main part of $\Iet_{\pi/2}(\Omega)+\Ito_{\pi/2}(\Omega)$ comes from the integrated signal.
In fact, for $\Omega =0$ we find from \eqref{eq:I1ex} that $\Iet_{\pi/2}(\Omega)/T=4/\gamma$ which even matches the QFI.
Considering two-time correlations we find for $\Omega\ll \gamma$, $F^{(1)}_{\pi/2}(\tau) \simeq -2\Omega^2 \e{-\gamma t/2}/\gamma,$
so by \eqref{eq:I2}, the Fisher information is
\begin{align}\label{eq:limit}
\frac{\mathcal{I}_{\pi/2}^{(2)}(\Omega)}{T} &\simeq  \frac{16\Omega^2}{\gamma^3}.
\end{align}
The Fisher information, \eqref{eq:I2} involves the square of the derivative of the correlation functions with respect to the sought parameter, so \eqref{eq:limit} hints, that the $n$-time correlation functions scale as $\Omega^{n}$, leading to contributions to the total classical Fisher information scaling as $\Omega^{2n-2}$.
To prove this relationship, we Taylor expand $e^{\mathcal{L}\tau}$ and $\rst$ to lowest non-vanishing order in $\Omega/\gamma$. Via \eqref{eq:multi} this reveals that for $\Omega/\gamma\ll1$, we indeed have $F_{\pi/2}^{(n/2)}(\{\tau_j\}) \propto \Omega^n$. Hence, we obtain for small $\Omega$ a series expansion,
\begin{align}\label{eq:OmegaExpansion}
\mathcal{I}_{\pi/2}(\Omega)&=a_0+\Omega^2 a_2+\Omega^4 a_4+\dots,
\end{align}
 where the $a_i$ are $\Omega$-independent.
This emphasizes that higher order correlations are less important at weak driving, but they, indeed, account for most of the information as the Rabi frequency is increased. This is illustrated in the high $\Omega$ regime of \fref{subfig:I}, where the information in $\Ito_{\pi/2}(\Omega)$ far from exhausts the QFI.

We find for
$\Omega \gg \gamma$ that
$
F^{(1)}_{\pi/2}(\tau) \simeq \gamma \cos (\Omega \tau) \e{-3\gamma \tau/4},
$
and by \eqref{eq:fisher} this leads to a Fisher information valid at strong driving,
\begin{align}
\frac{\mathcal{I}^{(2)}_{\pi/2}(\Omega)}{T} = \frac{256\Omega^2\left(243\gamma^4+216\gamma^2\Omega^2
+128\Omega^4\right)}{27\gamma\left(9\gamma^2+16\Omega^2\right)^3},
\end{align}
where taking the limit leads to
\begin{align}\label{eq:asymFisher}
\frac{\mathcal{I}^{(2)}_{\pi/2}(\Omega)}{T} = \frac{8}{27\gamma}\qquad\text{for }\quad \Omega\rightarrow \infty.
\end{align}
This is independent of $\Omega$, showing that once the transition is fully saturated, increasing the driving strength further does not alter the information available from the two-time correlation function.
This can be readily understood, since exactly the same information is available from the power spectrum \eqref{eq:FisherW}. Beyond saturation, the spectrum $\Sw$ in \fref{fig:F1Mollow} consists of two side peaks of unchanged shapes and separated by $2\Omega$. $\Omega$ can hence be determined with the same precision for all large values.
Conversely, the integrated signal holds negligible information beyond saturation,
where the steady state does not depend on the actual value of $\Omega$.

Finally, we note that $\Ito_{\pi/2}(\Omega)$ is maximized at $\Omega = 0.71\gamma$, where $\Iet_{\pi/2}(\Omega)$ is zero, so here including at least two-time correlations in the data record is essential.

\subsection{Finite detector efficiency}
\begin{figure}
\centering
\includegraphics[trim=0 0 0 0,width=0.9\columnwidth]{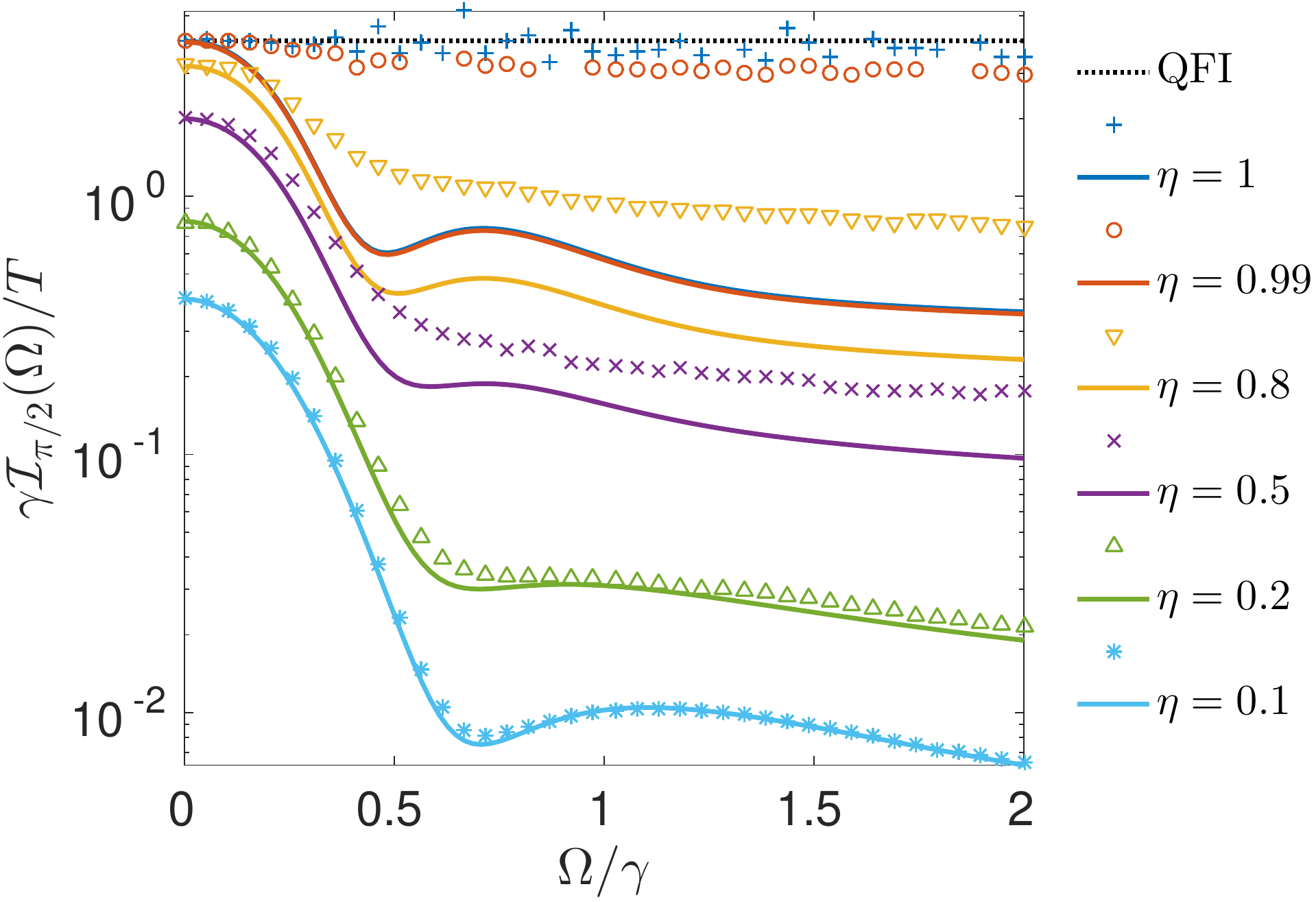}
\caption{(Color online)
The full classical Fisher information for homodyne detection (markers) is shown for different values of the detector efficiency $\eta$ and
compared to the combined information from the integrated signal and two-time correlations $\Iet_{\pi/2}(\Omega)+\Ito_{\pi/2}(\Omega)$ (lines in the same order). The results are shown as functions of the actual Rabi frequency $\Omega$ for probing with the local oscillator phase $\Phi = \pi/2$.
\label{fig:FisherEta}
}

\end{figure}

Finally, we address the distribution of information when the detector is imperfect ($\eta<1$).
While under perfect detection, the conditional system state in \eqref{eq:1}, remains pure, this is not the case when $\eta<1$. This is a consequence of our inability to trace the state exactly when some photo emission events are missed by the detectors, and inevitably leads to a decreased information in the signal.

The integrated current \eqref{eq:I} scales as $\sqrt{\eta}$ which by \eqref{eq:I1} is reflected as an $\eta$-scaling in the information from this part.
In general, it follows from \eqref{eq:J} and \eqref{eq:multi} that
\begin{align}\label{eq:FnScaling}
F^{(n/2)}_\Phi(\{\tau_j\})\propto \eta^{n/2},
\end{align}
so that
\begin{align}\label{eq:etaScaling}
\mathcal{I}^{(n)}_\Phi(\theta)\propto \eta^n.
\end{align}
Hence, higher order correlations in the signal are less important as the detection efficiency decreases and correspondingly by \eqref{eq:etaScaling}, the information in the signal is fully contained in the lowest order temporal correlations, i.e. for low $\eta$ we expect $\mathcal{I}_\Phi(\theta) \simeq \mathcal{I}^{(1)}_\Phi(\theta) +\mathcal{I}^{(2)}_\Phi(\theta)$. As seen in
\fref{fig:FisherEta}, where we compare $\mathcal{I}^{(1)}_{\pi/2}(\theta) +\mathcal{I}^{(2)}_{\pi/2}(\theta)$ to the Fisher information of the full homodyne current \eqref{eq:FI} for different values of $\eta$ as a function of the Rabi frequency, this is indeed the case. In particular, for realistic detection efficiency $\eta\lesssim 0.5$ the full information is largely contained in the lowest order time-correlations of the signal, and with low efficiency detection  $\eta\lesssim 0.1$ higher order correlations hold only negligible information as seen by the light blue data in the figure.

\section{Conclusion and outlook}
We have calculated the classical Fisher information for homodyne detection of the radiative emission from a resonantly driven two-level system and demonstrated that a Bayesian signal analysis reaches the corresponding Cramér-Rao sensitivity bound on the applied Rabi drive strength. The classical Fisher information of homodyne detection is upper bounded by the quantum Fisher information associated with the quantum state of the (unmeasured) field emitted by the system, and lower bounded by the information retrieved by the mean value of the homodyne signal.

The emitter system is subject to measurement backaction due to the noisy measurement data, and the signal thus acquires non-trivial temporal correlations, which are at the heart of the performance of the Bayesian analysis. In the case of photon counting, the waiting time distribution between subsequent detection events, and hence the two-time intensity-intensity correlation function accounts for the full data record, and suffices to compute the Fisher information and explain its dependence on the physical parameters. Here we have investigated the information contained within the two-time correlation function or, equivalently, the power spectrum of the homodyne detection current. Due to the weaker measurement backaction, homodyne detection outcomes may be correlated for a larger number of different measurement times and, in particular for strong driving, we have observed that most of the information about the Rabi frequency is hidden in multi-time correlations in the current and can be retrieved efficiently only by the Bayesian analysis.

Our methods of analysis are general and apply to detection of signals from any quantum system and for the estimation of any physical parameter governing the evolution of that system. There have been theories suggesting that a sufficiently weak measurement back action might lead to an asymptotic resolution scaling better than $1/\sqrt{T}$ \cite{PhysRevLett.114.210801}, and that more complex systems will offer similar improvement \cite{PhysRevA.93.022103,arXiv:1604.06400}. Our two-level example system does not offer long memory times and, despite the weak homodyne probing, it is subject to the ergodicity arguments given in Sec. I, implying  a $1/\sqrt{T}$ scaling of sensitivity. The QFI and the theory for the classical Fisher information, however, constitute a good starting point for a general investigation of the parameter resolution limit offered by quantum systems.

\section{Acknowledgements}
The authors acknowledge financial support from the Villum Foundation.



%

\end{document}